# How Does Environmental Information Disclosure Affect Corporate Environmental Performance? – Evidence from Chinese A-Share Listed Companies

Zehao Lin

The Ohio State University, department of economics

1734 Summit, Columbus, OH 43201, USA

Zehao8461@gmail.com

**Abstract**

Global climate warming and air pollution pose severe threats to economic development and public safety, presenting significant challenges to sustainable development worldwide. Corporations, as key players in resource utilization and emissions, have drawn increasing attention from policymakers, researchers, and the public regarding their environmental strategies and practices. This study employs a two-way fixed effects panel model to examine the impact of environmental information disclosure on corporate environmental performance, its regional heterogeneity, and the underlying mechanisms. The results demonstrate that environmental information disclosure significantly improves corporate environmental performance, with the effect being more pronounced in areas of high population density and limited green space. These findings provide empirical evidence supporting the role of environmental information disclosure as a critical tool for improving corporate environmental practices. The study highlights the importance of targeted, region-specific policies to maximize the effectiveness of disclosure, offering valuable insights for promoting sustainable development through enhanced corporate transparency.

## 1 Introduction

Global climate warming and air pollution pose serious threats to national economies and public health. Extreme weather events and environmental degradation strain infrastructure, reduce productivity, and result in significant economic losses. Additionally, air pollution severely impacts respiratory health, contributing to a wider range of chronic conditions, such as asthma, bronchitis, other lung diseases, eventually lowering quality of life. All these hinder sustainable development initiatives. Corporations, as major sources of greenhouse gas emissions and pollutants, have long been recognized as an important agent for mitigating global climate warming and air pollution. As global concerns about climate change and environmental protection deepen, the responsibility of corporations to promote sustainable development and reduce emissions is becoming increasingly important. Numerous studies have focused on the environmental performance of corporations, primarily focuses on three key areas: corporate environment performance measurement (Trumpp, Endrikat, & Zopf, 2015), determinants of corporate environmental performance (Arco-Castro et al., 2024), and its effect on national economies (Chin et al., 2024). From the perspective of determinants, area-level factors such as temperature (Wang & Ogawa, 2015), precipitation (Amil et al., 2016), industrial structure (Dong, Sun, Li, et al., 2024), and government intervention (Li & Qi, 2024) have been found to significantly affect corporate environmental performance. At the firm level, variables such as company size and technological innovation also play critical roles in influencing corporate environmental performance (Schäfer et al., 2020; Deng et al., 2022).

In recent years, environmental information disclosure has emerged as a potential strategy for mitigating air pollution caused by corp. Theoretically, greater transparency in environmental information disclosure increases corporate accountability and encourages firms to prioritize environmental responsibility, thereby adopting environmental protection measures. However, Amel-Zadeh and Serafeim (2018) highlight that some companies allocate excessive resources to preparing and publishing environmental information disclosure materials while neglecting substantive improvements. In such cases, environmental performance may stagnate or even deteriorate due to a disproportionate focus on reporting over actual remediation efforts. Consequently, the impact of environmental information disclosure on corporate environmental performance remains an area warranting further investigation. Existing studies have largely overlooked regional heterogeneity in the impact of environmental information disclosure and corporate environmental performance. Moreover, the relationship between environ inf dis and corporate environmental performance remains predominantly theoretically, with limited empirical evidence.

To address these gaps, this study examines the impact of environmental information disclosure on corporate environmental performance among A-share listed companies in China. Using a panel two-way fixed effects model, we analyze how environmental information disclosure influences corporate environment performance and assess the heterogeneity of these effects across regions with varying population sizes, per capita technology expenditure, and green coverage rate. Furthermore, we investigate the underlying mechanisms driving these effects.

The contributions of this study are threefold. First, it enriches the existing empirical studies on the impact of corporate environmental information disclosure on corporate environment performance. Second, it highlights the regional heterogeneity of these effects, shedding light on how different contextual factors mediate the relationship. Finally, it provides an in-depth analysis of the mechanisms underlying the observed effects, offering insights into how environmental information disclosure translates into environmental outcomes.

The remainder of this paper is organized as follows. Section 2 provides a literature review. Section 3 discusses the research methodology and data sources. Section 4 presents the empirical results and their analysis. Finally, Section 5 concludes with a discussion of the findings and their implications.

**2.1 Factors Influencing Environmental Performance**

Natural factors play an important role in shaping environmental performance. Climatic conditions, such as temperature, precipitation, and wind speed, significantly influence environmental impacts (Amil et al., 2016; Hien et al., 2002). High temperatures and wind speed up the decomposition and dispersion of pollutants, while precipitation plays a critical role in dissolving and diluting pollutants. The influence of climatic conditions on pollutant reductions has been evidenced in several countries, including China and Vietnam (Zhao et al., 2020; Zhang et al., 2017; Hien et al., 2002). Topography plays a significant role in affecting environmental performance. In areas with flat topography, air can move freely, helping to disperse pollutants and reduce smog concentration. However, regions with complex topography, such as terrains, valleys, and mountains, restrict air movement, facilitating the accumulation of pollutants and increasing air pollution (Giovannini et al., 2020).

The impact of economic growth on environmental performance is controversial. Economic growth is usually accompanied by large amount of increases energy consumption and industrial pollutant emissions, degrading air quality and straining environmental systems. For example, studies on Chinese cities link rapid economic growth and industrialization to worsening pollution levels (Shaw et al., 2010; Canh et al., 2021). Moreover, the pressures from urbanization and population growth amplify resource exploitation, further escalating environmental degradation. However, economic growth enables investments in clean technologies, renewable energy, and sustainable infrastructure, reducing reliance on pollution-intensive activities. Sustainable infrastructure development has demonstrated long-term economic and environmental benefits (Mahmood et al., 2024). For instance, China's focus on green finance has significantly supported renewable energy projects, enhancing corporate and regional sustainability. Additionally, economies transitioning to higher-value industries with lower pollution intensities can achieve better environmental outcomes. Green investment and reduced energy intensity are critical to balancing growth with sustainability (Ullah et al., 2024). In addition, the Environmental Kuznets Curve (EKC) hypothesis suggests that pollution rises with early economic development but declines after reaching a certain income level. (Shaw et al., 2010). This suggests that economic growth may also have a nonlinear impact on environmental pollution.

Pollution Haven Hypothesis emphasizes that foreign direct investment (FDI) can deteriorate environmental quality in host countries, especially developing ones. This is because corporations from developed countries transfer high-pollution factories to countries with lax environmental regulations. For instance, central and western China have experienced higher pollution levels due to lenient environmental oversight, as FDI attracts pollution-intensive industries, thereby worsening air quality (Wang & Liu, 2019). Conversely, the pollution halo effect shows that FDI can improve environmental quality through technology spillover. This is because FDI brings advanced, cleaner production technologies. When paired with stringent environmental standards, FDI can encourage cleaner production practices and technology transfers, reducing firms' pollution intensity by enhancing productivity and environmental management capabilities, as observed in regions with stronger regulations (Wang & Liu, 2024). These contrasting effects underscore the dual-edged impact of FDI on environmental performance and the need for robust policy frameworks to maximize its environmental benefits.

Infrastructure development and technological advancements are essential for improving environmental performance. For example, in China, expanding public transportation systems has significantly reduced private vehicle usage and improved air quality by reducing total pollutant emissions (Qiu & He, 2017). However, some scholars point out that traffic-related emissions remain a primary source of air pollution, particularly in regions like Beijing–Tianjin–Hebei and Chengdu–Chongqing (Wang et al., 2022). Technological progress is often lined to improved environmental quality. Advancements in tech can drive cleaner production methods, minimize pollutant emissions at their source, enhance resource efficiency, and further reduce environmental pollution.

The role of government intervention in shaping environmental performance has been a critical focus of recent research. Among various initiatives, the Government Environmental Information Regulation (GEIR) stands out as a significant driver of environmental performance, particularly in the environmental and social dimensions, achieved through improved information disclosure and green innovation efficiency (Li et al., 2024). This effect is especially evident in firms with low political relevance, high investor attention, and regions with low marketization, underscoring the importance of targeted and well-designed regulatory efforts to promote corporate sustainability (Li et al., 2024).The effect of government intervention on environmental performance is also complex and multifaced. Government intervention, on the hand, can improve environmental performance through implementing an enforcing environmental standards, emissions limits and pollution control. On the other hand, certain government environmental policies may impose short-term constraints on environmental performance, because local government may prioritize economics growth over environmental protection, leading to increase carbon emission and pollution. For instance, the Natural Resource Asset Departure Audit pilot program in China showed a significant negative impact on the environmental performance of companies, especially among non-state-owned enterprises and in specific regions. This suggests that while the overarching goal of such policies is to enhance environmental sustainability, poorly designed or implemented measures can increase operational pressures, potentially hindering environmental performance improvements (Yan et al., 2023)

## 2.2 The Impact of Environmental Information Disclosure on Corporate Environmental Performance

Based on existing research, corporate environmental information disclosure can influence corporate environmental performance through several positive pathways. Firstly, environmental information disclosure can improve regional environmental management by enhancing pollutant treatment rates and local government regulatory efforts, which subsequently improves regional environmental conditions and enhances corporate environmental performance (Wang, Shang, Li, & Li, 2023). Secondly, disclosing environmental information can enhance a company's reputation and public image, thereby attracting investor attention, particularly green finance support. The funds obtained can be used for technological innovation, improving environmental technology levels, and strengthening a company's environmental governance capabilities (Reid & Toffel, 2009). Thirdly, environmental information disclosure can increase corporate transparency, attract investors focused on sustainability, and further promote corporate investment in environmental technology research and development, which facilitates technological upgrades and positively impacts environmental governance (Lyon & Maxwell, 2011). Fourthly, environmental information disclosure encourages companies to focus on supply chain management, promoting the development of green supply chains, which reduces the environmental burden across the industry chain and achieves more efficient resource utilization (Downar et al., 2021).

However, some scholars have raised concerns regarding the role of environmental information disclosure. On the one hand, environmental information disclosure can lead to increased management costs, especially in the short term. This is particularly challenging for small enterprises, which may cut back on environmental protection expenditures in response to disclosure requirements, thus negatively affecting their environmental performance (Wang, Shang, Li, & Li, 2023). On the other hand, some companies may engage in selective disclosure or exaggerate their environmental actions, causing the disclosed information to fail in effectively guiding the allocation of social resources, resulting in actual negative impacts on environmental improvement (Lyon & Maxwell, 2011). Therefore, while environmental information disclosure contributes to improving corporate environmental performance, it is important to address issues related to cost and authenticity in its implementation to fully realize its utility.

## 2.3 Gaps and Research Framework

A comprehensive review of existing literature reveals several gaps in research on corporate environmental performance: First, most studies focus on either firm-level or regional-level factors, lacking an integrated analytical framework that combines both dimensions. Second, research exploring the relationship between environmental information disclosure (E) and corporate performance often stops at examining the overall correlation, with limited attention paid to the heterogeneity of this relationship across different contexts. Third, while much of the literature discusses the theoretical mechanisms through which E influences corporate environmental performance, there is a lack of empirical validation of these mechanisms.

To address these gaps, we construct an analytical framework (Figure 1) that integrates both firm-level and regional-level factors. The core focus of this study is the impact of environmental information disclosure on corporate environmental performance, analyzed through a comprehensive framework that incorporates both dimensions. We hypothesize that corporate and regional factors interact to shape corporate environmental performance outcomes, and as such, we include the following control variables: At the firm level, we consider the number of employees (ten thousand people), average technological expenditure per capita (n 10,000 CNY), total debt ratio, and proportion of tertiary industry in regional GDP to reflect the internal characteristics and operational context of firms. At the regional level, we include variables such as GDP per capita (in CNY), population density, green space coverage, government budgetary activity (in CNY), trade openness (in 10000 USD) and internet penetration to account for the broader economic, infrastructural, and social environment in which firms operate.

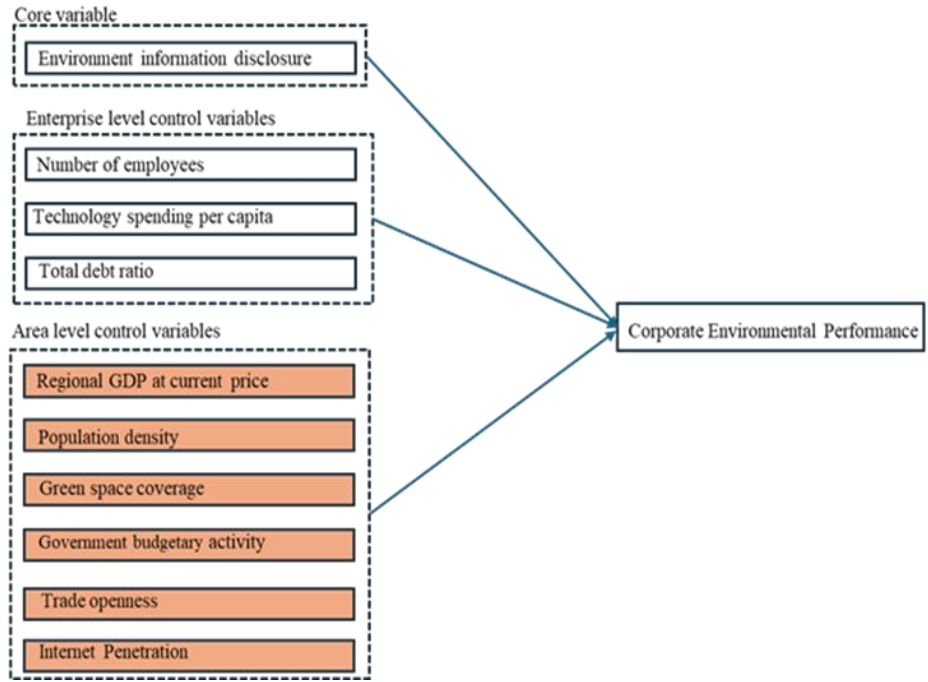

## 3 Data and Method

### 3.2 Research Method

This study used panel two-way fixed effect model to estimate the effect of environmental information disclosure on corporate environmental performance. The formula used was as following:

$$CEP = \theta_i + \alpha E + \beta X_{it} + \gamma_t + \varepsilon_{it}$$

Where corporate performance represents the performance of corporate entity i in period t. The core explanatory variable is E, indicating environmental information disclosure. α is the corresponding coefficient. $X_{it}$ was the control variable vector a range of factor that could potentially affect Corporate environmental performance. $\theta_i$, $\gamma_t$, $\varepsilon_{it}$ represented fixed effects, time effects, and residual, respectively. The descriptive statistical analysis of each variable is presented in Table 1.

Table 1 Description statistic of variables

| Control Variables | Shortened Variables | Description | Source of Variables |
|---|---|---|---|
| PM2.5 | PM2.5 | Fine particulate matter smaller than 2.5 micrometers, harmful to air quality and health. | Self-generated grid data was used to match the PM2.5 concentration at each enterprise's location |

| Environment Information Disclosure | E | Publicly sharing data on environmental impact and sustainability. | GTA Database |
|---|---|---|---|
| Total Population | TP | Total number of residents in a given region. | China City Statistical Yearbook |
| Number of Employees | NOE | Total number of employees in an organization or company. | GTA Database |
| Per Capita Technology Expenditure | PCTE | Total Science and Technology Expenditure divided by Total Population | China City Statistical Yearbook |
| Total Debt Ratio | TDR | The proportion of total debt to total assets, indicating financial risk | GTA Database |
| Proportion of Tertiary Industry in Regional GDP | POTIIRGDP | The share of the service sector in regional GDP, reflecting its economic importance. | China City Statistical Yearbook |
| Per Capita Regional GDP | PCRGDP | The regional GDP divided by the total population | China City Statistical Yearbook |
| Population Density | PD | Number of people per square kilometer of land area. | China City Statistical Yearbook |
| Green Space Ratio | GSR | The proportion of green space to the total land area in a region, reflecting environmental greening levels. | China City Statistical Yearbook |
| Government Budget Ratio | GBR | The ratio of government revenue and expenditure to regional GDP, indicating fiscal involvement. | China City Statistical Yearbook |
| Foreign Trade Dependency | FTD | The share of actual foreign investment in regional GDP, showing reliance on international trade. | China City Statistical Yearbook |
| Internet Penetration Rate | IPR | The proportion of broadband internet users to the total population, measuring digital accessibility. | China City Statistical Yearbook |

In this paper, $X_{it}$ include several control variables: PCTE, POTIIRGDP, PCRGDP , PD, GSR, GBR, and FTD. Previous studies suggest that per capita technological expenditure may have a positive impact on corporate environmental performance (Zhao et al., 2022). This is because higher PCTE often reflects a region's focus on innovation and sustainable development, which can provide firms with advanced technologies and resources to improve their environmental performance. The proportion of POTIIRGDP might be positively correlated with corporate environmental performance (Wang & He, 2024); at the regional level, PCRGDP could have a favorable effect on corporate environmental performance (Huang et al., 2022); while PD appears to be negatively associated with corporate environmental performance (Chen & Zhang, 2023). Moreover, green space coverage is considered to potentially have a positive influence on corporate environmental performance (Li et al., 2022); GBR may promote corporate environmental performance to some extent (Kim & Park, 2023); and in regions with insufficient industrial restructuring, FTD might negatively impact corporate environmental performance (Xu & Li, 2023). The specific effects of these variables require further research and validation to better understand their mechanisms and contextual dependencies.

**4 Examination Method**

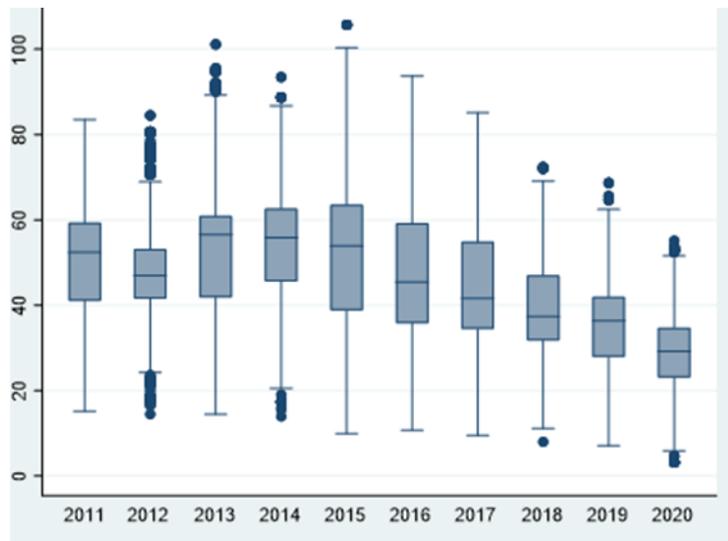

Table 2 Time Dynamics of Corporate Environmental Performance

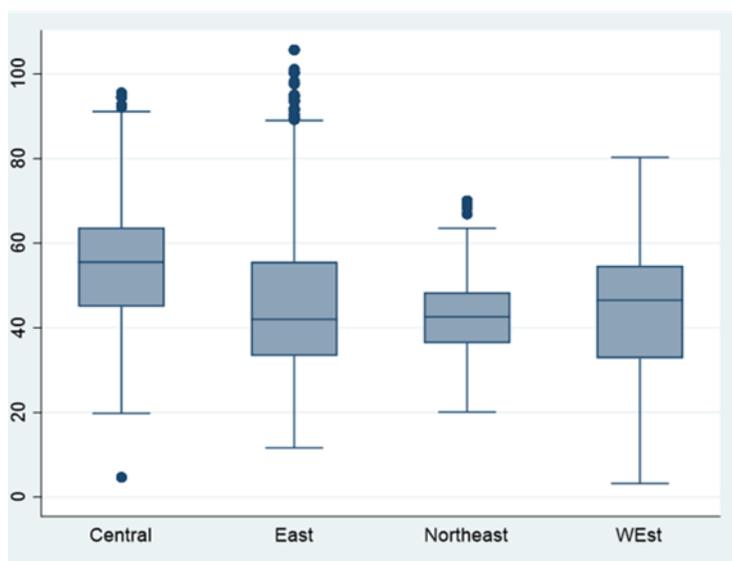

Table 3 Corporate Environmental Performance in the Context of China's Economic Zones

The spatial evolution of corporate environmental performance demonstrates a pattern of stabilization followed by rapid improvement, as evidenced by various empirical studies. This positive shift has been particularly notable since 2015, highlighting significant advancements in China's corporate environmental practices in recent years, especially beginning in 2015. In 2015, SynTao Green Finance introduced the Environmental, Social, and Governance index, which serves as a benchmark for assessing whether a corporation is operating sustainably and contributing positively to society. The Environmental Information Disclosure component included in the Environmental, Social, and Governance index provides the public with opportunities to evaluate the environmental impact of corporations. This, in turn, encourages corporations to enhance their production methods and improve their environmental performance, ultimately earning greater public recognition. Moreover, the observed improvements in corporate environmental performance can also be attributed to the implementation of stringent environmental policies by the Chinese government during this period. These policies likely played a pivotal role in shaping corporate behavior and driving improvements in environmental outcomes. However, the observed decline in environmental impact is a complex phenomenon, influenced by a combination of corporate efforts and policy interventions. Therefore, further empirical analysis is necessary to disentangle and quantify the effects of these factors.

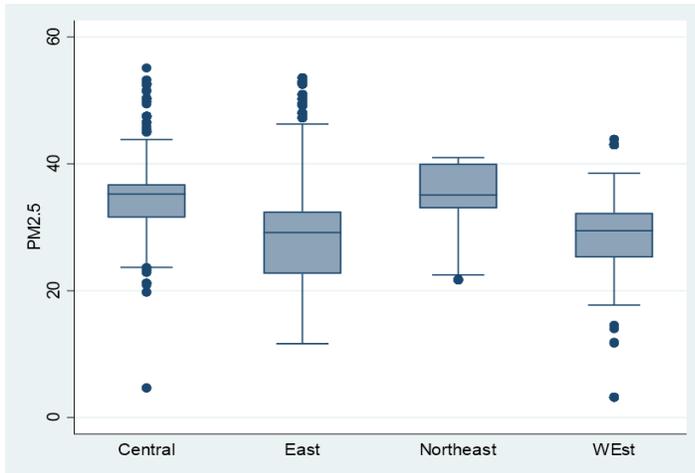

Table 4 Corporate Environmental Performance Across China's Economic Zones in 2011

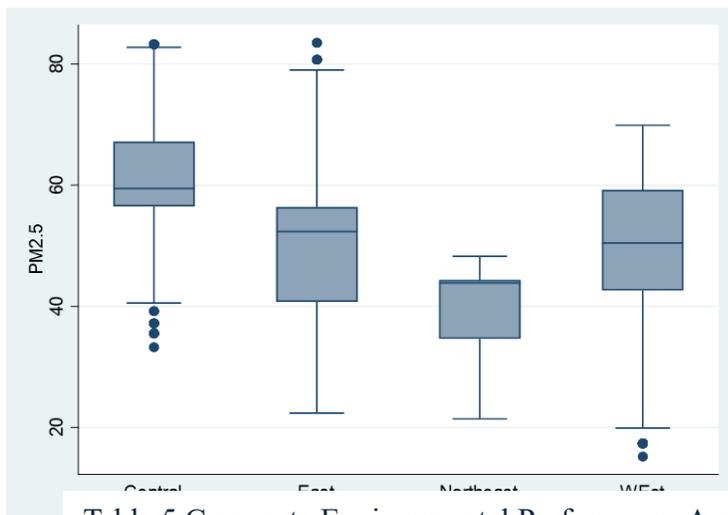

Table 5 Corporate Environmental Performance Across China's Economic Zones in 2020

From the perspective of spatial distribution, there are significant regional disparities in corporate environmental performance across the East, Central, West, and Northeast regions of China. Corporations in the Eastern region exhibit the highest corporate environmental performance, likely due to their advanced technological capabilities and strong environmental governance. This is followed by the Central region, where moderate industrial pollution is offset by emerging technological advancements. The Western region, characterized by its vast geographical expanse and a slightly higher median concentration of pollutants, faces challenges stemming from its reliance on heavy industries and relatively underdeveloped environmental technologies. Meanwhile, the Northeast region, despite being an old industrial base, shows relatively lower median PM2.5 concentrations. This can be attributed to slower economic growth and the decline of traditional industries, which has led to reduced emissions. However, the Northeast region may experience localized pollution spikes during the winter heating season, potentially linked to the widespread use of coal for heating. This seasonal phenomenon highlights the need for targeted interventions to address persistent environmental challenges in this region.

Over time, regional corporate environmental performance has shown significant trends of change, potentially influenced by various factors such as regional economic restructuring, technological advancements, and resource utilization. In 2011, the eastern region demonstrated relatively strong environmental performance, likely due to its advanced technological capacity, which mitigated the environmental pressures of intensive economic activities. The western region followed, possibly benefiting from its lower population density, sparse economic activity, and fewer industrial pollution sources. Despite its relatively low technological development, these natural advantages helped maintain a favorable environmental performance. In contrast, the central and northeastern regions showed relatively weaker performance. The central region, heavily reliant on resource-based industries, particularly coal mining in provinces such as Shanxi, faced significant environmental pressure. Limited technological capabilities further constrained pollution control efforts. Similarly, the northeastern region, as a traditional industrial base, struggled with a high proportion of heavy industries and delayed industrial transformation, contributing to its lower environmental performance.

By 2020, the environmental performance landscape had shifted noticeably across regions. The eastern region maintained its leading position, likely supported by continued technological advancements, strong policy backing, and sustained economic optimization. The northeastern region saw improvements in environmental performance, potentially reflecting the effects of industrial restructuring and enhanced pollution control measures. Technological advancements also likely began to play a larger role in mitigating environmental issues. The western region exhibited stable performance, benefiting from its low population density and limited pollution sources. While economic activities in the west were increasing, the overall pollution levels remained relatively low. The central region, although showing some improvement, continued to lag behind, with minimal progress in narrowing the performance gap. This may be attributed to its reliance on resource-based industries such as coal mining, which exert significant environmental pressure. The high population density in some areas further exacerbated environmental challenges, and the region's technological capacity remained below that of the east and northeast.

**4.1 Association between E and PM2.5**

We apply a two-way fixed effects panel model to examine the relationship between E and PM2.5 concentration, as shown in module 1. Initially, only E is included in the model in Model 1, we find that E is significantly and negatively related to PM2.5. Considering that the existence of omitted variables in the model can lead to biased estimation, we add a series of factors that may affect PM2.5, to minimize the potential estimation bias. After controlling these variables, E is still significantly and negatively related to PM2.5. The results indicate that the improved environmental information disclosure may reduce PM2.5 concentration. Environmental information disclosure allows the public and non-governmental organizations to understand the pollution emissions of enterprises. By increasing external oversight, it forces enterprises to adopt cleaner production technologies and reduce pollution emissions.

Among the control variables, several factors, including green space coverage, industrial structure, and trade openness, show significant effects on PM2.5 concentration. Specifically, the coefficient for GSR is negative, indicating that higher green coverage may help absorb pollutants, thereby reducing PM2.5 levels (Luo, Chen, Sheng, & Wang, 2023). The coefficient for the proportion of POTIIRGDP is positive, suggesting that the growth of the service sector could increase emissions. For example, studies have found that consumption activities in the tertiary industry drive significant pollution emissions along supply chains, especially in interprovincial trade, where consumer provinces may shift pollution emissions to producer provinces (Wang, Zhang, Liu, Li, & Liu, 2024). Additionally, FTD trade openness may lead to increased energy consumption and pollution through industrial expansion, potentially raising PM2.5 levels (Cole, M. A., Elliott, R. J., & Zhang, L., 2017). Incorporating these control variables enhances the explanatory power of the model and provides insights into the potential impacts of E on PM2.5 concentration.

Table 6 The impact of environmental performance on corporate environmental performance

(1)

r1

| VARIABLES | pm25 |
|---|---|
| e | -0.0515** |
| | (0.0230) |
| TP | 0.00303 |
| | (0.00338) |
| NOE | 2.01e-05* |
| | (1.13e-05) |
| PCTE | -0.000944*** |
| | (7.47e-05) |
| TDR | 0.937 |
| | (1.040) |
| POTIIRGDP | 0.000305*** |
| | (4.45e-05) |
| PCRG | -9.66e-06*** |
| | (1.74e-06) |
| PD | -0.166 |
| | (0.163) |
| GSR | -0.125*** |
| | (0.0220) |
| GBR | -8.35e-06*** |
| | (1.25e-06) |
| FTD | 1.29e-05*** |
| | (1.51e-06) |
| IPR | -0.000266** |
| | (0.000107) |
| Year dummy | Control |
| Constant | 39.28*** |
| | (2.837) |
| | |
| Observations | 6,007 |
| Number of symbol | 1,015 |
| R-squared | 0.614 |

Robust standard errors in parentheses
*** $p<0.01$, ** $p<0.05$, * $p<0.1$

### 4.2 Robust test

We conducted two robustness tests to demonstrate the reliability of our results. First, to address potential estimation bias caused by extreme outliers, we applied truncation tests at the 1% and 5% levels (Models 1 and 2 in Table 7). Second, we used total employment to measure firm size.

Notably, firm size can also be measured using fixed assets, as both total employment and fixed assets reflect the scale of a firm's resources and production capacity. Total employment represents the scale of human resources and reflects labor input, while fixed assets capture capital investment and production infrastructure, representing the physical resources and operational capacity of a firm. To avoid potential estimation bias arising from different methods of measuring firm size, we replaced total employment with fixed assets (Model 3). Furthermore, we replaced the share of the secondary industry in GDP with the share of the tertiary industry, as both indicators can effectively reflect the industrial structure (Model 4). Additionally, we replaced total population with registered population because registered population more accurately reflects the number of long-term residents within a region, excluding the impact of transient populations. This improves data stability and explanatory power, particularly for analyzing the long-term relationship between local economic activities and environmental outcomes (Model 5).

Table 7 Robustness tests on the impact of E on corporate environmental performance

| VARIABLES | (1) Model1 | (2) Model2 | (3) Model3 | (4) Model4 | (5) Model5 |
|---|---|---|---|---|---|
| e | -0.0421** | -0.0489** | -0.0480** | -0.0479** | -0.0519** |
|   | (0.0201) | (0.0213) | (0.0231) | (0.0230) | (0.0230) |
| TP | -0.000892 | 0.00443 | 0.00325 | 0.00265 |   |
|   | (0.00401) | (0.00362) | (0.00339) | (0.00323) |   |
| NOE | 1.43e-05 | 1.85e-05* |   | 1.75e-05 | 2.21e-05* |
|   | (1.06e-05) | (1.08e-05) |   | (1.10e-05) | (1.15e-05) |
| PCTE | -0.00103*** | -0.000999*** | -0.000930*** | -0.00101*** | -0.000961*** |
|   | (7.33e-05) | (7.73e-05) | (7.42e-05) | (7.88e-05) | (7.36e-05) |
| TDR | 0.882 | 0.977 | 1.038 | 0.969 | 0.773 |
|   | (0.952) | (1.007) | (1.045) | (1.075) | (1.039) |
| POTIIRGDP | 0.000154*** | 0.000310*** | 0.000306*** |   | 0.000222*** |
|   | (4.49e-05) | (4.49e-05) | (4.45e-05) |   | (4.37e-05) |
| PCRGDP | -1.26e-05*** | -7.59e-06*** | -9.68e-06*** | -7.26e-06*** | -1.11e-05*** |
|   | (1.80e-06) | (1.68e-06) | (1.74e-06) | (1.86e-06) | (1.67e-06) |
| PD | 0.685*** | 0.107 | -0.169 | 0.273* | 0.267* |
|   | (0.129) | (0.133) | (0.163) | (0.149) | (0.161) |
| GSR | -0.0821*** | -0.112*** | -0.125*** | -0.125*** | -0.0873*** |
|   | (0.0181) | (0.0216) | (0.0221) | (0.0217) | (0.0194) |
| GBR | -5.92e-06*** | -8.84e-06*** | -8.39e-06*** | -2.10e-06*** | -4.69e-06*** |
|   | (1.22e-06) | (1.29e-06) | (1.25e-06) | (7.27e-07) | (1.21e-06) |
| FTD | 5.82e-06*** | 1.37e-05*** | 1.30e-05*** | 5.15e-06*** | 1.07e-05*** |
|   | (1.66e-06) | (1.56e-06) | (1.51e-06) | (8.65e-07) | (1.39e-06) |
| IPR | -0.000315*** | -0.000283*** | -0.000266** | -0.000194* | -0.000329*** |
|   | (8.78e-05) | (0.000107) | (0.000107) | (0.000115) | (9.87e-05) |
| Total Size |   |   | 0* |   |   |
|   |   |   | (0) |   |   |
| Proportion of Secondary Industry in GDP |   |   |   | 0.220*** |   |
|   |   |   |   | (0.0603) |   |
| Registered Population |   |   |   |   | -0.000870*** |

|  |  |  |  |  | (0.000109) |
|---|---|---|---|---|---|
| Year Dummy | Control | Control | Control | Control | Control |
| Constant | 40.09*** | 37.01*** | 39.30*** | 30.67*** | 41.58*** |
|  | (3.315) | (3.043) | (2.851) | (3.416) | (0.765) |
| Observations | 5,328 | 5,865 | 6,010 | 6,005 | 6,007 |
| Number of symbol | 1,008 | 1,015 | 1,015 | 1,015 | 1,015 |
| R-squared | 0.597. | 0.629. | 0.613. | 0.603. | . 0.621 |

Robust standard errors in parentheses
*** p<0.01, ** p<0.05, * p<0.1

## 4.3 Endogeneity Test

Endogeneity arises from two primary sources: omitted variable bias and reverse causality. The inclusion of a comprehensive set of control variables in the model has partially mitigated the issue of omitted variables. Reverse causality, on the other hand, pertains to the bidirectional relationship between environmental information disclosure (E) and corporate environmental pollution levels. Specifically, firms with higher levels of E may be incentivized to enhance their environmental technologies, thereby reducing pollutant emissions. Conversely, firms with lower levels of pollutant emissions may be more inclined to disclose environmental information to gain goodwill from governments and the public, as well as to attract additional investment.

To address this endogeneity, we implemented an instrumental variable (IV) regression analysis. Drawing on prior research (Casey & Klemp, 2017) we employed historical levels of E as an instrumental variable for current E. The relevance of this instrument is rooted in its ability to influence current E practices through historical disclosure levels, which subsequently affect PM levels. The exogeneity of the instrument is supported by the assumption that current PM levels do not exert a causal influence on historical E practices.

The first-stage regression results confirm the validity of the instrument, with the Kleibergen-Paap Wald rk F statistic reaching 1025.40, surpassing the critical threshold for a 15% maximal IV size. Additionally, the Cragg-Donald Wald F statistic of 3848.16 further substantiates the strength of the instrument.

In the second stage, the results corroborate the effectiveness of the instrument. The Kleibergen-Paap Wald rk F statistic and Cragg-Donald Wald F statistic remain consistent at 1025.40 and 3848.16, respectively, underscoring the robustness of the first-stage instrument. Notably, the coefficient of E remains negative and statistically significant, providing compelling evidence of its role in reducing PM levels. These findings reinforce the robustness of the model and establish a credible causal relationship between E and pollutant emissions.

Table 8 Heterogeneity analysis of the impact of E on corporate performance

| VARIABLES | (1) pm25 |
|---|---|
| e | -0.0630** |
|  | (0.0279) |
| TP    | 0.00589*     |
|       | (0.00327)    |
| NOE   | 2.47e-05**   |
|       | (1.13e-05)   |
| PCTE  | -0.000923*** |
|       | (0.000101)   |
| TDR   | 1.184        |
|       | (0.937)      |
| POTIIRGDP | 0.000279*** |
|       | (3.94e-05)   |
| PCRGDP | -7.47e-06*** |
|       | (1.92e-06)   |
| PD    | 0.108        |
|       | (0.179)      |
| GSR   | -0.0989***   |
|       | (0.0276)     |
| GBR   | -8.68e-06*** |
|       | (1.15e-06)   |
| FTD   | 1.32e-05***  |
|       | (1.42e-06)   |
| IPR   | -0.00852**   |
|       | (0.00371)    |
| Year Dummy | Control |
| Observations | 5,070   |
| Number of symbol | 969 |
| R-squared | 0.623    |
| cdf   | 3874         |

Robust standard errors in parentheses
*** p<0.01, ** p<0.05, * p<0.1

## 4.4 Heterogeneity Test (Per Capita Science and Technology Expenditure, Total Population, Greening Rate)

Considering the effect of E on corporate environmental performance may vary across region with population size, technological investment level and green space coverage, we conduct the heterogeneity test. The findings indicate that in regions with higher population densities, the positive association between E and corporate environmental performance is more pronounced. This suggests that densely populated areas may amplify the effectiveness of E, likely due to greater public scrutiny, heightened environmental awareness, and increased pressure on firms to comply with environmental standards. The interaction between E and per capita technological expenditure reveals a significant heterogeneity. In regions with lower levels of technological investment, E has a stronger positive impact on corporate environmental performance. This is because in these regions, limited by technology level, companies may have previous addressed basic environment issues. Thus, implementing relatively simple environmental measures can significantly improve performance by reducing emissions, optimizing resource utilization, and enhancing energy efficiency, ultimately contributing to sustainable development and improved corporate reputation. In contrast, in regions with higher technological expenditure, firms may have already leveraged advanced technologies to enhance their environmental performance, thereby reducing the incremental effectiveness of E. A similar pattern emerges in the interaction between E and green space coverage. While green spaces independently contribute to improved environmental quality, their interaction with E in regions with extensive green space coverage appears to moderate the direct effect of E on corporate environmental performance. This attenuation may be attributed to the pre-existing ecological advantages in such regions, which limit the additional improvements that E can achieve. Conversely, in regions with limited green space, E plays a more critical role in enhancing firms' environmental practices by compensating for the lack of natural ecological benefits.

Table 9 Mechanism analysis

| VARIABLES | (1) pm25 | (2) pm25 | (3) pm25 |
|---|---|---|---|
| e | 0.0876** | -0.103*** | -0.0725*** |
|  | (0.0421) | (0.0281) | (0.0255) |
| e*TP | -0.000167*** |  |  |
|  | (3.54e-05) |  |  |
| e*PCTE |  | 2.45e-05*** |  |
|  |  | (4.38e-06) |  |
| e*GSR |  |  | 0.00182*** |
|  |  |  | (0.000594) |
| Control variable | Control | Control | Control |
| Year dummy | Control | Control | Control |
| Constant | 36.80*** | 41.06*** | 39.99*** |
|  | (2.799) | (2.822) | (2.827) |
| Observations | 6,007 | 6,007 | 6,007 |
| Number of symbol | 1,015 | 1,015 | 1,015 |
| R-squared | 0.616 | 0.616 | 0.615 |

Robust standard errors in parentheses
*** $p<0.01$, ** $p<0.05$, * $p<0.1$

**5 Conclusion**

Against the backdrop of global warming and air pollution threatening human survival and sustainable development, this study investigates the impact of environmental information disclosure (E) on corporate environmental performance and presents several key findings. First, environmental performance exhibits distinct temporal and spatial trends. Over time, corporate environmental performance follows a pattern of initial stabilization followed by rapid improvement. Spatially, the Eastern region demonstrates the highest performance due to its advanced technologies and robust environmental governance, while the Central and Western regions show moderate improvements, and the Northeast faces unique challenges, including its reliance on heavy industries and seasonal pollution spikes. Second, baseline regression and robustness checks confirm that E significantly enhances environmental performance by improving regional air quality, demonstrating that greater transparency in corporate environmental practices effectively reduces pollutant emissions. Third, the effects of E exhibit notable regional heterogeneity, with stronger impacts observed in areas of high population density and significant technological investment. Additionally, regions with greater green space coverage experience amplified positive effects of E. Fourth, mechanism analysis reveals that E indirectly improves environmental performance by optimizing corporate resource allocation, such as increasing total assets, and enhancing financial structures, including reducing financing costs.

These findings suggest several important policy implications. Governments should actively encourage enterprises to enhance the quality and transparency of environmental information disclosure, leveraging it as an effective tool to improve air quality and corporate environmental performance. Policy frameworks must be tailored to local conditions, leveraging the technological and demographic advantages of developed regions while providing targeted support for less-developed areas to avoid a one-size-fits-all approach. Additionally, efforts to optimize corporate resource allocation and improve financial structures, such as promoting green financing and strengthening corporate governance, can further amplify the benefits of environmental transparency. Integrating these measures with regional environmental strategies can enhance the overall impact of E across diverse regions.

Despite its contributions, this study has several limitations. First, the focus on A-share listed companies in China may limit the generalizability of the findings to other regions or unlisted firms. Future research could expand the sample to include companies from various sectors and regions, enabling more comprehensive analysis and refining policy insights. Second, while the study identifies the positive effects of E, further investigation is needed to assess the long-term sustainability of these improvements and potential unintended consequences. Third, although the study explores some mechanisms through which E influences environmental performance, additional pathways remain unexplored. Future research should examine a broader range of mechanisms, such as the role of green innovation and supply chain dynamics, to deepen understanding. Expanding the scope of analysis to include cross-regional comparisons and industry-specific characteristics will help refine these insights and provide a more holistic understanding of how E drives corporate and environmental sustainability.

# Reference

Amil, N., Latif, M. T., Khan, M. F., & Mohamad, M. (2016). Seasonal variability of PM2.5 composition and sources in the Klang Valley urban-industrial environment. *Atmospheric Chemistry and Physics, 16(8)*, 5357–5381. https://doi.org/10.5194/acp-16-5357-2016

Arco-Castro, M. L., López-Pérez, M. V., Alonso-Conde, A. B., & Suárez, J. R. (2024). Determinants of corporate environmental performance and the moderating effect of economic crises. *Baltic Journal of Management, 19(6)*, 39–59. https://doi.org/10.1108/bjm-06-2023-0233

Chin, W.C., Rasiah, R. & Furuoka, F. (2024). The corporate environmental and financial performance nexus: a comparison of corporations in China and Japan. *Energy Efficiency, 17*, 29. https://doi.org/10.1007/s12053-024-10212-1

Cole, M. A., Elliott, R. J., & Zhang, L. (2017). Foreign direct investment and the environment. *Annual Review of Environment and Resources, 42(1)*, 465–487. https://doi.org/10.1146/annurev-environ-102016-060916

Dong, Y., Sun, Z., Li, W. et al. (2024). How the low-carbon city pilot policy reduces enterprise carbon emissions and paves the way for high-quality development: evidence from a quasi-natural experiment in China. *Environment, Development and Sustainability*. https://doi.org/10.1007/s10668-024-05724-w

Downar, B., Ernstberger, J., Reichelstein, S. et al. (2021). The impact of carbon disclosure mandates on emissions and financial operating performance. *Review of Accounting Studies, 26*, 1137–1175. https://doi.org/10.1007/s11142-021-09611-x

Giovannini, L., Ferrero, E., Karl, T., Rotach, M. W., Staquet, C., Castelli, S. T., & Zardi, D. (2020). Atmospheric Pollutant Dispersion over Complex Terrain: Challenges and Needs for Improving Air Quality Measurements and Modeling. Atmosphere, 11(6), 646. https://doi.org/10.3390/atmos11060646

Li, C., & Qi, L. (2024). Can Government Environmental Attention Improve Corporate Carbon Emission Reduction Performance? —Evidence from China A-Share Listed Companies with High-Energy-Consumption. *Sustainability, 16(11)*, 4660. https://doi.org/10.3390/su16114660

Luo, S., Chen, W., Sheng, Z., & Wang, P. (2023). The impact of urban green space landscape on PM2.5 in the central urban area of Nanchang city, China. *Atmospheric Pollution Research, 14(11)*, 101903. https://doi.org/10.1016/j.apr.2023.101903

Lyon, T. P., & Maxwell, J. W. (2011). Greenwash: Corporate environmental disclosure under threat of audit. *Journal of Economics & Management Strategy, 20(1)*, 3-41. https://doi.org/10.1111/j.1530-9134.2010.00282.x

Li, X., Hu, Y., Guo, X., & Wang, M. (2024b). Government Environmental Information Regulation and Corporate ESG performance. Sustainability, 16(18), 8190. https://doi.org/10.3390/su16188190

Mahmood, S., Misra, P., Sun, H. et al. (2024). Sustainable infrastructure, energy projects, and economic growth: mediating role of sustainable supply chain management. *Annals of Operations Research*. https://doi.org/10.1007/s10479-023-05777-6

Park, K., Yoon, T., Shim, C., Kang, E., Hong, Y., & Lee, Y. (2020). Beyond Strict Regulations to Achieve Environmental and Economic Health-An Optimal PM2.5 Mitigation Policy for Korea. *International Journal of Environmental Research and Public Health, 17(16)*, 5725. https://doi.org/10.3390/ijerph17165725

Shaw, D., Pang, A., Lin, CC. et al. (2010). Economic growth and air quality in China. *Environmental Economics and Policy Studies, 12*, 79–96. https://doi.org/10.1007/s10018-010-0166-5

Shu, Z., Zhao, T., Liu, Y., Zhang, L., Ma, X., Kuang, X., Li, Y., Huo, Z., Ding, Q., Sun, X., & Shen, L. (2022b). Impact of deep basin terrain on PM2.5 distribution and its seasonality over the Sichuan Basin, Southwest China. *Environmental Pollution, 300*, 118944. https://doi.org/10.1016/j.envpol.2022.118944

The influence of firm size on the ESG Score: Corporate Sustainability Ratings under review on JSTOR. (n.d.). www.jstor.org. https://www.jstor.org/stable/45386660

Ullah, O., Zeb, A., Shuhai, N. et al. (2024). Exploring the role of green investment, energy intensity and economic complexity in balancing the relationship between growth and environmental degradation. *Clean Technologies and Environmental Policy*. https://doi.org/10.1007/s10098-024-02953-5

Wang, H., Liu, H. (2019). Foreign direct investment, environmental regulation, and environmental pollution: an empirical study based on threshold effects for different Chinese regions. *Environmental Science and Pollution Research, 26*, 5394–5409. https://doi.org/10.1007/s11356-018-3969-8

Wang, J., & Ogawa, S. (2015). Effects of meteorological conditions on PM2.5 concentrations in Nagasaki, Japan. *International Journal of Environmental Research and Public Health, 12(8)*, 9089–9101. https://doi.org/10.3390/ijerph120809089

Wang, J., Zhang, H., Liu, Y., Li, Z., & Liu, Z. (2024). Unexpected PM2.5-related emissions and accompanying environmental-economic inequalities driven by "clean" tertiary industry in China. *The Science of the Total Environment, 919*, 170823. https://doi.org/10.1016/j.scitotenv.2024.170823

Wang, L., Shang, Y., Li, S., & Li, C. (2023). Environmental Information Disclosure-Environmental Costs Nexus: Evidence from Heavy Pollution Industry in China. *Sustainability, 15(3)*, 2701. https://doi.org/10.3390/su15032701

Wang, W., Yang, S., Yin, K., Zhao, Z., Ying, N., & Fan, J. (2022). Network approach reveals the spatiotemporal influence of traffic on air pollution under COVID-19. *Chaos: An Interdisciplinary Journal of Nonlinear Science, 32(4)*. https://doi.org/10.1063/5.0087844

Yan, Y., Cheng, Q., Huang, M., Lin, Q., & Lin, W. (2022). Government Environmental Regulation and Corporate ESG Performance: Evidence from Natural Resource Accountability Audits in China. International Journal of Environmental Research and Public Health, 20(1), 447. https://doi.org/10.3390/ijerph20010447Zhang, B., Jiao, L., Xu, G. et al. (2018). Influences of wind and precipitation on different-sized particulate matter concentrations (PM2.5, PM10, PM2.5–10). *Meteorology and Atmospheric Physics, 130*, 383–392. https://doi.org/10.1007/s00703-017-0526-9

Zhang, L., Guo, X., Zhao, T., Xu, X., Zheng, X., Li, Y., Luo, L., Gui, K., Zheng, Y., & Shu, Z. (2022). Effect of large topography on atmospheric environment in Sichuan Basin: A climate analysis based on changes in atmospheric visibility. *Frontiers in Earth Science, 10*. https://doi.org/10.3389/feart.2022.997586

Zhao, X., Sun, Y., Zhao, C., & Jiang, H. (2020). Impact of Precipitation with Different Intensity on PM2.5 over Typical Regions of China. *Atmosphere, 11(9)*, 906. https://doi.org/10.3390/atmos11090906